\documentclass[a4paper, 11pt, final]{IEEEtran}
\usepackage[utf8]{inputenc}
\usepackage{graphicx}
\usepackage[pdftex,
            pdfauthor={Chemnitz, Bonnet, Shklovski, Buttrich, Watts},
            pdftitle={Unionized Data Governance in Virtual Power Plants},
            pdfsubject={},
            pdfproducer={Latex with hyperref},
            pdfcreator={pdflatex}]{hyperref}
\hypersetup{
  colorlinks   = true, 
  urlcolor     = blue, 
  linkcolor    = blue, 
  citecolor    = blue  
}
\usepackage[style=ieee,sorting=none,url=false,isbn=false,doi=false,eprint=false]{biblatex} 
\bibliography{DataGovernance}

\title{Unionized Data Governance in Virtual Power Plants}
\author{
  \IEEEauthorblockN{
    Niels Ørbæk Chemnitz\IEEEauthorrefmark{1},
    Philippe Bonnet\IEEEauthorrefmark{1},
    Irina Shklovski\IEEEauthorrefmark{2},
    Sebastian Büttrich\IEEEauthorrefmark{1} and Laura Watts\IEEEauthorrefmark{3}
    }\\
  \IEEEauthorblockA{
    \IEEEauthorrefmark{1}The IT University of Copenhagen, \texttt{\{niec,phbo,sbut\}@itu.dk}\\
  }
  \IEEEauthorblockA{
    \IEEEauthorrefmark{1}University of Copenhagen, \texttt{ias@di.ku.dk}\\
  }
  \IEEEauthorblockA{
    \IEEEauthorrefmark{2}The University of Edinburgh, \texttt{l.watts@ed.ac.uk}
  }
}

\begin{document}

\maketitle
\begin{abstract}
Flexible electricity networks continuously coordinate and optimize operations through ICT systems. An overlay data grid conveys information about the state of the electricity grid, 
as well as the status of demand and production of electricity in households and industry. Data is thus the basis for decisions that affect electricity costs and availability 
of assets on the electricity grid. It is crucial that these decisions are formed and monitored according to a well-defined governance model. No such data governance model exist today.
In this paper, we focus on the central role of virtual power plants in flexible electricity networks. We define the problem of data governance in a virtual power plant, insisting 
on the issues linked to the inherent asymmetry of this system. We propose unionization as a framing device to reason about this issue. The central contribution of this paper is thus
principles for a unionized data governance model for virtual power plants.
\end{abstract}

\section{Introduction}

Electricity networks have become pivotal in our societies attempt to 
decrease the emission of greenhouse gasses. 
Flexible electricity networks must constantly mediate the tension between the demand-agnostic production of 
electricity from renewable sources
and the supplier-agnostic consumption we have been accustomed to.
Flexible electricity networks continuously coordinate and optimize operations through ICT systems. 
An overlay data grid conveys information about the state of the electricity grid, 
as well as the status of demand and production of electricity in households and industry. 
Data is thus the basis for decisions that affect electricity costs and availability 
of assets on the electricity grid. It is crucial that these decisions are formed and monitored according to a well-defined governance model. No such data governance model exists today.
They are the topic of this paper.

We focus on the role of "Virtual Power Plants". Briefly for now, a virtual power plant, or VPP,
organizes the flows of data between many asset owners (individuals or companies) and an aggregator that takes decisions about tariffs and asset availability.
Since VPPs are a proposed way to integrate renewable energy sources and decrease the reliance on fossil fuels, they, simply by their function, inherit an aura of progressiveness, 
sustainability and collaboration. 
But the powerful position they hold, collecting sensitive personal data about individuals and remotely controlling assets in their households, opens many possibilities for exploitation. These must be counteracted by a strong governance model.

In this paper, we explore the issues around governance of the data that flows from individuals to aggregator within a VPP. We introduce the frame of 
unionization as a way to address these issues, and discuss how unions can be incorporated into data governance models to increase transparency, accountability and trust.

In Section II, we detail the role of virtual power plants in flexible electricity networks. Section III defines the problem in terms of data sources and data flows within the system. In Section IV, we introduce the frame of unionization as a way to address these problems, and Section V describes the principles for data governance derived from this frame. Section VI presents selected work that we find relevant for this issue, and Section VII summarizes the findings of the paper.

\section{Virtual Power Plants}

In this section we introduce the concept of Virtual Power Plants. We describe the motivation for flexibility in electrical networks, the inherently asymmetrical structure of VPPs, and the issues surrounding personal data collection.

\subsection{The Need for Flexibility}
Many energy-intensive sectors, including transportation and heating, are increasingly moving towards electrification. 
But the green promise of electrification can only be kept if the electrons flowing in our grids are from low-emission renewable energy sources. 
And while renewable generation capacity is growing, this is in itself not sufficient. 

Renewable energy sources --- such as wind, solar and tidal --- are \emph{non-dispatchable}, meaning that we cannot control their production. 
We can turn generators off, but we cannot generate electricity from these sources at will. The conditions of our environment must allow it. 
Sometimes, the wind is not blowing, the sun is not shining and the tidal currents are still. But our consumption of electricity continues regardless.
This creates large gaps in time between production and load, which is an issue as there is currently no viable solutions for large-scale energy storage.\footnote{
  Pumped-storage hydroelectricity is the primary large-scale energy storage solution, but it needs specific natural features (such as reservoirs) to be applicable.
} 
The electrical grid operators constantly balance the production and the consumption of energy in order to maintain a stable grid. 
So even if we generate enough renewable energy to match our demand on average, the need to balance consumption and non-dispatchable generation on a minute-to-minute basis is an on-going problem.

A proposed way to balance a grid with a large amount of non-dispatchable generation is to incorporate \emph{demand-side flexibility}~\cite{mohlerChapter23Energy2017}. 
The load on the grid is constrained due to physical limits of the infrastructure, so flexibility, or supply-following load, eases this constraint by managing the electrical demand 
– making demand dispatchable to balance the uneven renewable generation and its storage. 
Temporary energy storage and other types of dynamic demand response are crucial to managing renewable energy flexibly on the grid. 

But in order to achieve this flexibility, the future electricity networks, as envisioned by especially \cite{tanejaFlexibleLoadsFuture2013a} and \cite{katzInformationcentricEnergyInfrastructure2011}, will 
need to be digitally networked electrical grids that continuously coordinate and optimize operation through ICT systems (see also \cite{bondyFunctionalReferenceArchitecture2015, heArchitectureLocalEnergy2008,
amanPredictionModelsDynamic2015,sianoDemandResponseSmart2014}). This data infrastructure is a layer on top of the electrical grid, transporting information about the state of both industrial and domestic demand and 
generation. It organizes flows of data concerning home and individual energy use. These flows of data can be used to make decisions on how to balance the grid or on how to store electricity: Data is Electricity!

The data infrastructure associated with the electricity system relies on personal data. For example, home usage data in a 1-hour resolution can disclose whether 
you are home \cite{liisbergHiddenMarkovModels2016}, and sub-second resolution can reveal which movies you are watching \cite{grevelerMultimediaContentIdentification2012}. Data governance is essential 
to manage the impact of such data infrastructures on society.

A proposed component of this flexibly-managed smart grid, is what has popularly been coined \emph{Virtual Power Plants}\footnote{
  First proposed in \cite{veseyVirtualUtility1997} as "virtual utilities"
} (VPPs). A VPP provides flexibility by aggregating a host of smaller \emph{distributed energy resources} (DERs) --- such as solar panels, domestic batteries, electric vehicles or electric heating systems --- using their combined capacity to provide services normally reserved for industrial scale power plants.

\subsection{VPP Asymmetry}

The idea of VPPs is an emerging concept that has yet to see complete consensus on its definition. In \cite{adu-kankamCollaborativeVirtualPower2018} a literature review is carried out which identified a convergence towards the following definition:
\emph{"A VPP is a virtual entity involving multiple stakeholders and
comprising decentralized multi-site heterogeneous technologies, formed
by aggregating dispatchable and non dispatchable distributed energy
sources and energy storage systems, including electric vehicles and
controllable loads. It is supported by information and communication
technology to form the equivalent of a single virtual power plant with
capacity to manage and coordinate its operations, ensuring power and
information flow among its stakeholders in order to minimize generation
costs, maximize profits, and enhance participation in demand
response programs as well as trade within the electricity market."}
\cite[p.~281]{adu-kankamCollaborativeVirtualPower2018}

\begin{figure*}[tb]
  \begin{center}
    \includegraphics[width=0.8\textwidth]{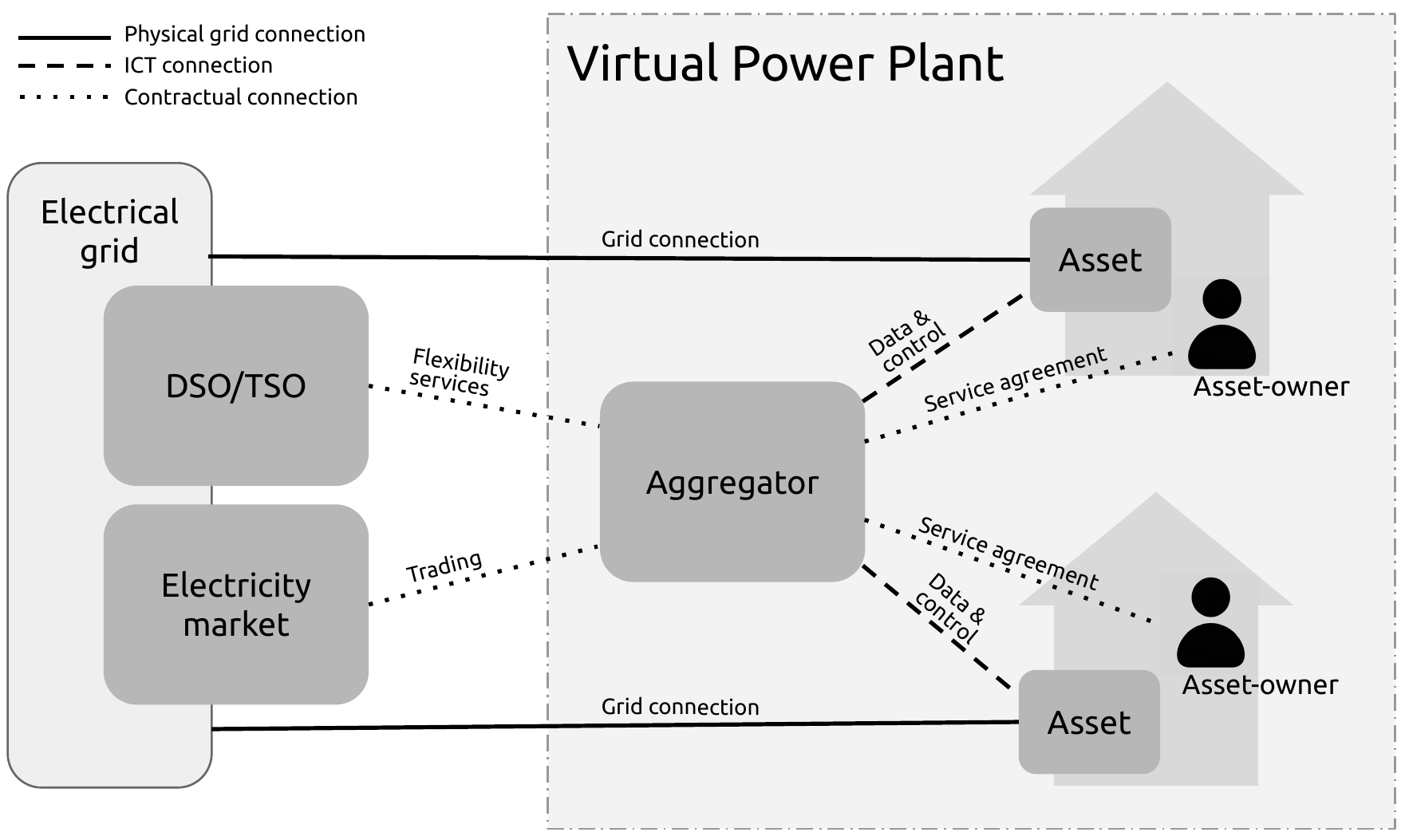}
    \caption{Overall structure of a VPP. The aggregator controls a portfolio of assets --- such as heat pumps, domestic batteries or EVs --- and participates in the energy market or provides flexibility services to power systems operators.}
    \label{fig:vpp-structure}
  \end{center}
\end{figure*}

The subset of VPPs of interest for this paper, are those in which the DERs, or \emph{assets}, are owned by individuals outside the organization that owns the VPP. The VPP as a concept includes all of the stakeholders, but we are particularly interested in relationship between the \emph{asset-owners} and the \emph{aggregator}. An asset-owner is a stakeholder that owns an individual asset in the portfolio of the aggregator. An example is a homeowner with a heat pump, electric car or domestic battery that can be remotely controlled. The aggregator\footnote{
  The term "aggregator" is sometimes used as a synonym for VPP, which is again a sign that the dust has not completely settled around the terminology in this field.
} is the stakeholder controlling the ICT system that collects data and controls the assets. The relationship between asset, aggregator and VPP is shown in \autoref{fig:vpp-structure}. 

\subsection{VPPs and Personal Data}

The ICT system at the core of the VPP is dependent on several data flows in order to function. The most interesting and complex from a governance perspective is the flow from asset-owner to aggregator. This data flow contains electricity consumption information from the home (and potentially electric car), which is highly personal data. Behavior patterns can be inferred from electricity consumption, and the data acts as the basis for decisions made about the control of the asset, which influences both power cost and availability for the asset-owner. Because of the nature and purpose of this data, the data governance model employed within the VPP is crucial for the development of trust between asset-owners and aggregator.

In a demonstration project conducted as part of the Danish \textsc{EcoGrid 2.0}-project \cite{heinrichEcoGridLargescaleField2020}, an aggregator was implemented, controlling heat pumps in 300 private households over the course of 3 years \cite{mullerLargescaleDemonstrationPrecise2019}. This aggregator remotely sent throttling commands to the heat pumps, decreasing load on the grid at certain times. The aggregator collected individual electricity consumption measurements every 5 minutes, and used it as input to its model along with weather service data. In the context of \textsc{EcoGrid 2.0} the homeowners were found to have a high degree of trust in the project, attributed in part to the technicians from the local community-owned utility, who installed and serviced the connected heat pumps \cite{pallesenArticulationWorkMiddle2018}.

The mass use of internet-connected services and devices that collect and store data of our use and behaviors have led to a continuous and pervasive monitoring of our lived lives. This data collection is generally not approved by the users, who find the data collection "creepy" \cite{shklovskiLeakinessCreepinessApp2014}. Still, despite regular public outcries, people continue to use the services, handing over their data. This has typically been described as an expression of the so-called "privacy paradox" \cite{norbergPrivacyParadoxPersonal2007,barnesPrivacyParadoxSocial2006}, which describes the discrepancy between intentions and behaviors when it comes to personal information disclosure. 

Recent research suggest that this "paradox" represents wide-spread "digital resignation" \cite{draperCorporateCultivationDigital2019a}: People desire to control their personal information, but feel unable to do so, rendering them passive. 
A study on data leakage in mobile apps conducted in \cite{shklovskiLeakinessCreepinessApp2014} indicated that people were opposed to this practice of data collection, but could see no alternative except to stop using the smartphone as a smartphone altogether.
The organizations collecting data benefit from this resignation, cultivating it through obfuscation measures, such as complicated privacy policies and transparency initiatives that allows partial access but not control \cite{draperCorporateCultivationDigital2019a}.  

What differentiates the case of VPPs from data collection through mobile apps is in part the novelty of the domain. Continuos monitoring of electrical consumption is a new type of surveillance in a primary private sphere, the home, and unlike the phones and smart speakers we bring into the home ourselves, this monitoring becomes part of the infrastructure that makes the house. Finally, the aggregator has direct control of the home, controlling electricity availability and cost. The aggregator is not just sensing, but also actuating.

\section{Problem Definition}

A VPP can be viewed as a system that collects data from various sources and combines these in order to make decisions about the behavior of its assets and, by consequence, the lives of the asset-owners. In order to define the problem of data governance between the asset-owners and aggregator, we must first define the different types of data sources and flows within the system. From these we identify the relationships that these flows represent, and the problems to address.

\subsection{Data Sources}

For an aggregator to function within a VPP, the minimum data needed is the current status of the assets. In addition to this, the aggregator needs the privileges to remotely control the assets, which is tightly coupled to data access. Status data from a single asset, such as battery charge level or heat pump temperature, becomes richer when collected over time, and can be used to model the behavior of residents. These models can estimate when assets are idle or homes are vacant, which is highly relevant information for a VPP. But the richness of this data also makes it valuable for many purposes outside the scope of a VPP, such as profiling with the purpose of targeting or manipulation.

Data from assets alone does not allow for much meaningful decision-making. The crucial property of VPPs is the ability to optimize for outside parameters. For commercial aggregators, this data might come from electricity markets, detailing the market prices of electricity now or in the future, allowing the aggregator to optimize for price and profit by buying power in bulk when it is cheap (e.g \cite{vayaOptimalBiddingStrategy2015, bessaOptimizedBiddingEV2012}). Or it might come from DSOs, detailing bids for flexibility services in certain parts of the grid. For VPPs where profit is not the only goal, other relevant data sources might be the current generation of renewable energy, if the goal is to increase utilization of renewable energy, or the weather, if the goal is to stock up on energy in preparation for cold or extreme weather.

\subsection{Data Flows}

\begin{figure}
  \begin{center}
    \includegraphics[width=\columnwidth]{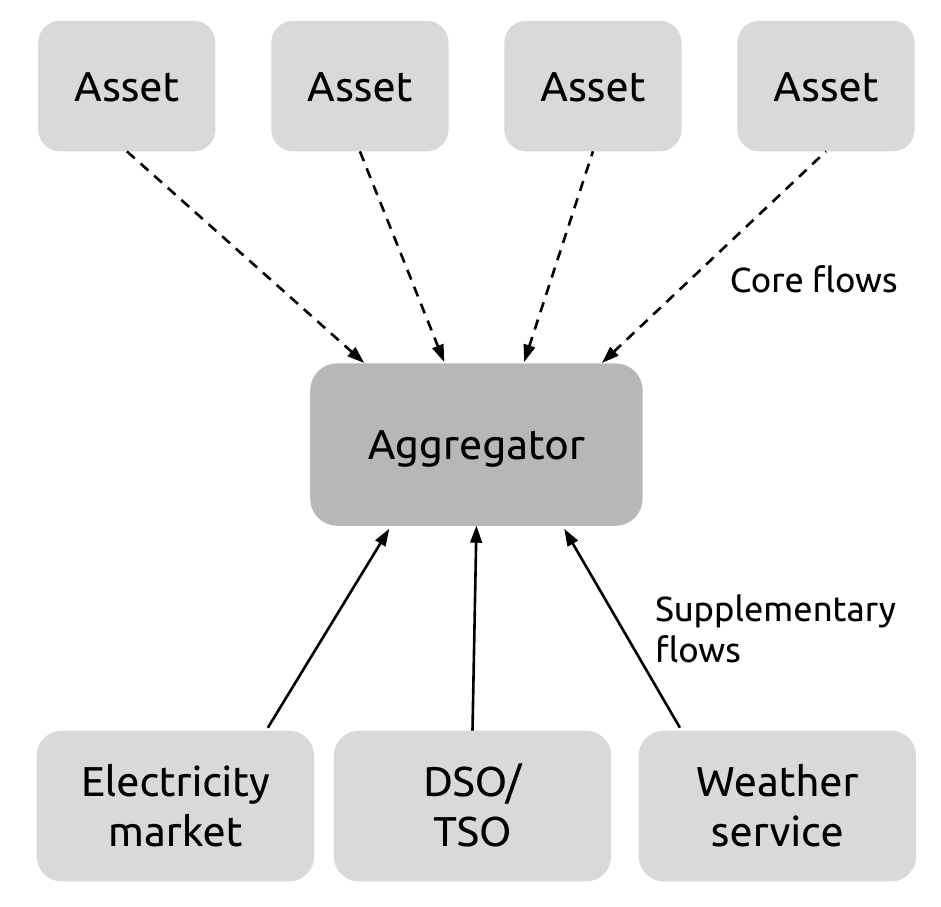}
    \caption{Data flows to the aggregator. The upper part represents the assets in the aggregators portfolio, while the lower part represents the data sources on which the aggregator bases its decisions.}
    \label{fig:data-flows-1}
  \end{center}
\end{figure}

The data created at each source flows to the aggregator to be used in its decision engine, deciding where to increase or decrease electricity consumption. We can distinguish between two discrete types of data flows: The \emph{core flows}, data which flows from the assets to the aggregator, and the \emph{supplementary flows}, which is data from external sources used in the optimization algorithm of the aggregator (see \autoref{fig:data-flows-1}).

\subsubsection{Core Flows}
The core flows are the basis of the capabilities of the VPP, carrying the current status, and thereby potential, of the portfolio. Each of the core flows represent the relationship between an organization (the aggregator) and an individual (the asset-owner). The core flows are numerous, a one-to-many relationship, and each of the core flows are in practice interchangeable. The aggregator simply needs enough capacity, it is less important exactly which assets that capacity is comprised of. If a core flow is discontinued or unavailable, the VPP can either continue with slightly decreased capacity or replace the flow with a similar asset.

\subsubsection{Supplementary Flows}
The supplementary flows are the basis of the direction of the VPP, carrying the current outside parameters that are used when optimizing for system goals. The supplementary data flows have the characteristic in common that they are between two organizations. They are one-to-one relationships that are not interchangeable. Each supplementary flow contains data to be used in the optimization loop of the aggregators decision engine. If a supplementary data flow is discontinued or unavailable, the VPP might not be able to make decisions, rendering it idle.

Both the two categories of data flows, the core and supplementary flows, are necessary for the VPP to function. But due to the interchangeability, the individual core flow is of lower importance than the individual supplementary flow. 

\subsection{Vulnerability and Conflicting Goals} \label{seq:conflicting-goals}

The nature of these data flows creates an asymmetrical power relation between the aggregator and the asset-owner. The aggregator has control over the electricity cost and availability for the asset-owners and access to valuable and personal data about them. This makes issues around data governance --- how the data is used in the decision engine of the aggregator, and to what degree the personal data is analyzed and shared --- important to the asset-owners. Simultaneously, the asset-owner is in a poor position to make demands or bargain with the aggregator, as the single data flow the asset-owner represents is not critical to the function of the VPP. Each individual asset-owner has very limited leverage in such a negotiation.

This becomes an issue when the aggregator and asset-owners have conflicting goals. The workings of a system such as a VPP are a manifestation of its \emph{function} or \emph{purpose} \cite{meadowsThinkingSystemsPrimer2008}, to borrow terms from systems theory, which can be seen as the high-level direction of the system. This direction is shaped by the structural properties of the system, the context it is placed in, and the \emph{goals} of the stakeholders. Goals in this context represents properties of the \emph{desired direction} of the system, and are unique for each stakeholder. Examples of goals for a VPP include maximizing profits, lowering prices for end-consumers, ensuring grid stability or transitioning to renewable energy. The goals of a stakeholder are often not explicitly stated and their internal compatibilities are complex. Goals $A$ and $B$ can be said to be "conflicting" if adjusting the direction of the system to address $A$ negatively impacts $B$.    

For example, the aggregator presumably has a goal of maximizing profits, and the asset-owners have a goal of preserving the integrity of their personal sphere. Given the issue of selling data about the behavior of asset-owners to third parties, these two goals are in conflict. However, the decision about whether to sell the data or not, is placed with the aggregator. 


If an asset-owner wishes to alter the direction of the VPP, they are in a precarious position to do so. This is due to the asymmetry of political power and leverage, and the consequence of such a conflict: If the asset-owner stops the cooperation with an aggregator, the asset-owner loses all of the perks of the service, while the aggregator only loses a fraction of its capacity.

It is this governance question, of how the asset-owners can influence the way their data is being utilized in a system where their leverage is limited, that is addressed in this paper.

\section{The Unionization Frame}

In this section we present the problem in the frame of \emph{unionization}, inspired by the consensus-seeking trade unions of Northern Europe. From this context we derive principles for data governance in the context of VPPs.

\subsection{Managing Power Asymmetries Through Unions}

The core data flows of a VPP can be seen as analogous to the relations of employment in classical capitalism, where the production is contingent on the pooling of a significant amount of labor. This labor is secured through individual contracts (or simply day-wages, meaning no contracts at all), which leaves the worker in an asymmetrical power relation to the employer similar to the relationship described in \autoref{seq:conflicting-goals}.

The primary ways of addressing the asymmetry between worker and employer has been government regulations and the forming of \emph{trade unions}. Trade unions mediate the flow of labor from workers to employer, and through this mediation gain leverage when negotiating wages and conditions. This is what is known as \emph{collective bargaining}. Should the employer refuse to meet the demands of the trade union representing the workers, they can invoke a \emph{strike}, removing the employers access to labor and thereby halting production. 

In a more general sense, unions can be seen as \emph{aggregators of power}. Whether in relation to labour, consumer rights, student conditions, or a host of other domains, the fundamental concept remains: By organizing, the interchangeability and replaceability of the individual relation is countered by collective action.

\begin{figure}
  \begin{center}
    \includegraphics[width=\columnwidth]{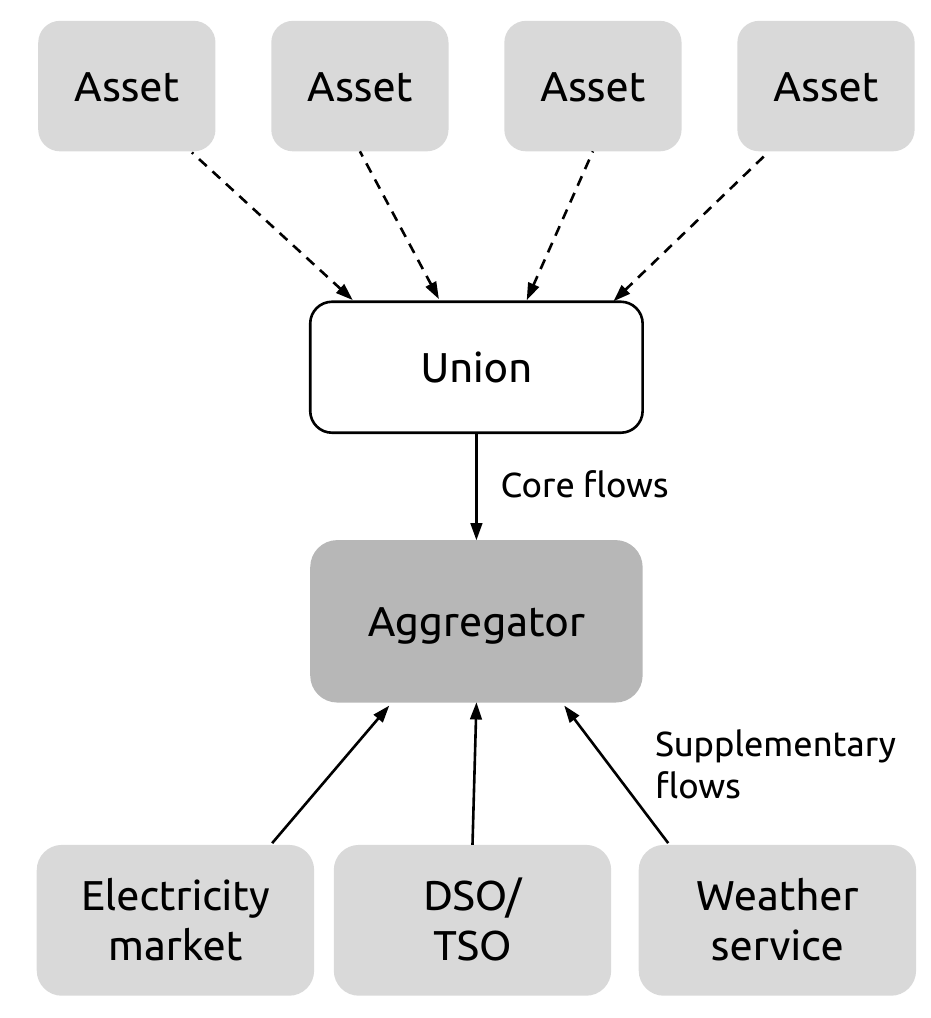}
    \caption{Unionized data flows to the aggregator. The core flows, the data from the assets, flows though a node of union-controlled infrastructure, allowing for leverage in negotiations with the aggregators}
    \label{fig:data-flows-2}
  \end{center}
\end{figure}

The main concepts from traditional trade unions can be used within a VPP as a way to obtain leverage and negotiating power in the relationship between aggregator and asset-owner. We propose the creation of a new institutional actor - the \emph{data union}, which individual asset owners can join. The data union has clearly articulated and defined goals and mediates on behalf of the asset owners, controlling all (or a significant amount of) the core data flows. This allows for collective bargaining, by using the possibility of halting or limiting the core flow (a "strike") as leverage.
The structure of unionized data flows is shown in \autoref{fig:data-flows-2}.

This mediation can be implemented in various ways, offering the union different degrees of power. The union could act like a traditional trade union, bargaining contract conditions on behalf on its members and, in case of a conflict, encouraging members to stop the data flow to the aggregator by turning of the connectivity of their collective assets. Since this union is interacting with an ICT system, there is also the possibility for much more direct and dynamic involvement of the union in the data flows. This could either be done by granting the union control over the assets in a way that trumps the aggregator, allowing them to dynamically start and stop the core flow. Alternatively the data flows could be rerouted to go through union-controlled infrastructure before being forwarded to the aggregator, creating a single point of control between the union and the aggregator.

If the union chooses to physically route the data through union-controlled infrastructure, this allows for a wider range of possibilities for control of data usage. One such possibility is anonymization through continuous de-identification. An aggregator needs to know which resources are available, but it does not strictly need to know which asset belongs to which asset-owner. By rehashing the identifiers of asset-owners at set intervals in time, the union can limit the aggregators ability to reason about the individual asset-owners behavioral patterns, while still allowing them access to all resource information. Controlling infrastructure also allows the union to construct a valuable dataset from which more general patterns and information can be deduced, something that has otherwise only been the privilege of the aggregator.

\subsection{Organizing Communities}

Traditionally, trade unions have been organized around fairly coherent occupational groups, whose members had similar conditions, experiences and demands as a base for organization. However, the diversification and specialization of the workforce have coincided with a great decline in union membership and power through the 1980s and 90s \cite{visserTradeUnionDecline2007, brysonIntroductionCausesConsequences2011, bellTradeUnionDecline1998, waddoupsTradeUnionDecline2005}. 

One way to address these issues has been to organize unions around \emph{"place-based communities"} other than the work\emph{place}, which can be seen by unions advocating for change on a social and political level that goes beyond worker self-interest \cite{holgateCommunityOrganisingUK2015}. As described in \cite{chaseEarlyTradeUnionism2017}, early trade unions were anchored in the local community, and \cite{levesqueLocalGlobalActivating2002} argues that a return to local community unions is "at the heart of union renewal strategies".

This harmonizes with the idea of a VPP data union. In contrast to many other asymmetrical relations of information, which can be highly globalized and detached from the physical spaces occupied by its users, VPPs are necessarily concentrated within a common locality, determined by the topology of the electrical grid. A VPP must have a significant amount of capacity in an area to offer services to the DSOs and compete on the local energy market. This means that the asset-owners have in common a relationship to that area, and the localized community could provide an arena to organize in.

Clearly, the hard borders of the grid topology will not reflect the fluid and dynamic borders of local communities, but the common relationship and frame of reference of the asset-owners, that the locality provides, can serve as the basis for the construction of a new community centered around electricity- and data-based issues. The data union would be a key part of this community.

\subsection{Enabling Productive Conflicts}

Conflicts between stakeholders can be seen as a steering mechanism for a system, altering its direction according to the goals of those involved. But in order for this mechanism to work, these conflicts have to be productive. And to enable productive conflicts, each side must acknowledge the power of the other. This creates an even foundation for the negotiations surrounding the conflict, taking into account the goals and interests of both sides. 


By unionizing, the asset-owners can force the aggregator to acknowledge their power, leveraging their ability for collective action. This can be done by halting or manipulating the flows of data or electricity to and from the assets. And while these collective actions are not without costs for the asset-owners, the mere ability to disrupt the functionality of the aggregator increases leverage.

\section{Governance Principles}

We have presented the issue of power asymmetry and exploitation within a VPP in the frame of unionization, and now precede to describe which principles for data governance that can be derived from this frame.

\begin{figure}
  \begin{center}
    \includegraphics[width=\columnwidth]{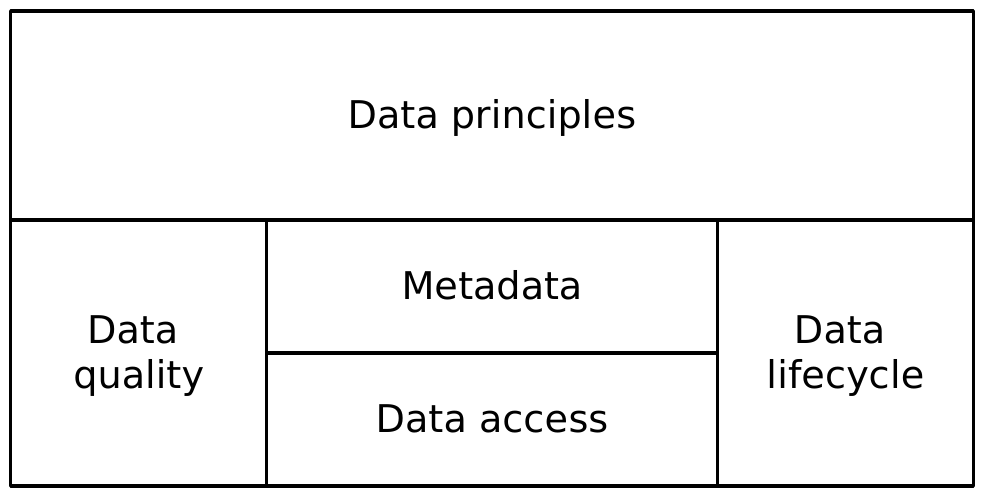}
    \caption{The five decision domains of data governance as defined in \cite{khatriDesigningDataGovernance2010}}
    \label{fig:domains}
  \end{center}
\end{figure}

To describe data governance models in general terms, we use the framework proposed by Khatri \& Brown, \cite{khatriDesigningDataGovernance2010}\footnote{
  Adapted from the IT governance model presented in \cite{weillITGovernanceHow2004}
} partitioning the issues of governance into five decision domains: \emph{data principles}, \emph{data quality}, \emph{metadata}, \emph{data access} and \emph{data lifecycle}. These domains influence each other, with the data principles informing the decisions and models in the remaining four domains. This relationship is visualized in \autoref{fig:domains}.

\subsection{Unionization of Data Subjects}

The basis for the following data governance principles, is the constitution of a data union that mediates the goals of the asset-owners.

By creating a union the asset-owners can engage in collective bargaining with the aggregator, negotiating the terms of the service that collects and processes information about their lives. The union replaces the one-to-many aggregator-to-asset-owner relationship with a one-to-one aggregator-to-union relationship. This gives the aggregator a single point of contact with the unionized asset-owners, and constitutes a single point of leverage, allowing asset-owners to make demands of the aggregator.

\subsection{Data Principles and Representation}

The decision domain of \emph{data principles} is concerned with "clarifying the role of data as an asset", its uses and purposes, and the communication of its uses. In other words, this is where the direction of all other decisions surrounding data governance is made. In the case of VPPs this includes whether selling or sharing data with third parties is part of the business model, whether modeling the behavior of individual assets should be allowed, which supplementary data sources the decision engine should be based on, and so on. 

These principles should define a clear business owner of data assets, and a governing body should be constituted to oversee the approvement and enforcement of data usage in the context of the defined purpose and direction. One example of such a governing body is an "Enterprise Data Committee". A committee of this sort would be an obvious place for union representation, allowing the interests and values of the asset-owners to be heard when discussing and reviewing data usage. It would provide an organizational frame where the union can impact the direction of the VPP by leveraging its control of the core data flow.

\subsection{Disruptive Measures for Asserting Power}

The data governance model should include concrete methods that allow for the asset-owners to disrupt the functioning of the VPP, if the current data usage does not live up to their demands. This could be done by limiting or halting the core flow of data, the VPPs basis for operation, thereby putting pressure on the aggregator to respond to their demands.

These measures are intended as a last resort, as none of the stakeholders are interested in a purely idle VPP. But its existence is important to establish the fact that the goals and demands of the asset-owners should be taken into account, in order to avoid damaging consequences for the aggregator.

This is in line with the Danish model for collective bargaining, where measures are taken to avoid disruptive conflicts such as a strike. This tradition is based on extensive negotiations towards a reconciliation that is acceptable to all parties. Only if the parties are unable to resolve the conflict, with government-assisted mediation, a strike is allowed. Still, these negotiations are always based on the fact that a strike is possible.

\subsection{Accountability Needs Transparency}

Khatri \& Brown describe the \emph{"locus of accountability of decision making"} in a decision domain to denote those held accountable if the handling of data does not live up to the data governance model. But before we can hold anybody accountable, we need to be able to assert that something has gone wrong. 

The aggregator and asset-owners might have conflicting goals, but the aggregator is responsible for the majority of the data collection, storage, processing, and usage. This means that the aggregator does not necessarily have an interest in divulging information about actions that diverges from the data principles. Therefore the asset-owners must implement measures to ensure a high degree of transparency, enabling them to audit the actions of the aggregator. 

From this we can see that transparency and auditing procedures around the implementation of the use of the data asset is crucial when designing a data governance model for VPPs.

\section{Related Work}

We present a couple of selected related works that offer other perspectives on similar or related issues of data and governance. We present the notion of "data commons", derived from commons theory, and "\textsc{Shift}", a future design project that addresses the asymmetrical relations of information within an existing trade union. 

\subsection{Data Commons}

Drawing on commons theory from the field of economic governance (e.g. \cite{ostromGoverningCommonsEvolution1990}), the frame of "data commons" seeks to describe the accumulated data of users as a \emph{common pool resource} (CPR). 

The main issue addressed in commons theory is how to ensure sustainable governance of a CPR that is \emph{rivalrous}, meaning that its use by one precludes its use by another, and where the locally optimal strategies for the individual agents leads to a suboptimal strategy when combined. A key example is fisheries, with the CPR being the stock of fish in the sea: Each fisher benefits more, in the short term, by catching as much as possible, but the combination of each fisher pursuing this strategy results in overfishing and eventual depletion and erosion of the CPR.\footnote{
  While often called the "tragedy of the commons", it is worth noting that more recent studies show how, in practice, it is not a preordained tragedy, but much richer, more ambiguous, what \cite{councilDramaCommons2002} describe as a "drama of the commons".
}

Data is unlike the traditional environmental CPRs, as it does not directly erode with overuse. Data can be copied and copied without loss. However, data and knowledge can be seen as rivalrous if value is not strictly measured as monetary worth, but instead described along three dimensions: \emph{economic}, \emph{sociological} and \emph{identity} (as described in \cite{aragonWhereCommonsMeet2011}). These dimensions become intuitively clearer if we think of sustainability as \emph{responsibility}: Irresponsible management is not confined to actions that diminish economic value, but also actions that affect sociological structures in a negative way or is at odds with the identities of the stakeholders. The primary solution to CPR issues is the creation of a locally anchored \emph{stewardship} role that oversees the usage of the resource.

A significant amount of work has been done applying commons theory to the non-depletable resources of \emph{knowledge} and \emph{information} (e.g. \cite{hessUnderstandingKnowledgeCommons2007, hessIdeasArtifactsFacilities2003}). 
In regard to data governance and commons theory, most research has been into the field of health care data (e.g. \cite{taylorWhatResponsibleSustainable2019,purtovaHealthDataCommon2017,evansBarbariansGateConsumerDriven2016}) and "smart cities" (e.g. \cite{beckwithDataFlowSmart2019}).

\subsection{\textsc{Shift} - A Platform Union}

In this paper, we have been looking at the unionization of data creation. The \textsc{Shift} project \cite{slavnovaSHIFT2017} is a future design project that takes the reverse approach. Based around OPR, the union of russian truck drivers, the \textsc{Shift} project proposes a strategic roadmap for the union, aiming to secure the living conditions of truckers as the development of automated solutions ("self-driving trucks") threatens to render them without employment.

To combat this threat, \textsc{Shift} plans to equip the trucks of their members with sensors (camera, GPS, accelerometer) and collect data from these in order to create a dataset that would be of great value when developing these automated solutions. The goal of this is to either use the dataset as leverage when bargaining with employers, or to simply circumvent the employers completely and use the dataset to develop union-owned automation, retrofitting the trucks of their members with autopilot systems, turning the union into a logistics company.

While speculative, this project has interesting parallels to the problem discussed in this paper, the primary being that the users own the assets, in this case the trucks. This gives them freedom to create, terminate, and redirect data flows to and from those assets. The strategic roadmap also includes the consolidation of the union through a digital platform, allowing for a sense of community despite the extremely distributed nature of truck driving. A similar solution might also be beneficiary in a VPP context for asset-owners to discuss internal issues.

\section{Conclusion}

In this article we have presented the frame of unionization, inspired by trade unions, as a way to describe the inherent problem of asymmetric relations of information in VPPs and its consequences for data governance models. We model the flows of data within a VPP as the \emph{core data flows} from assets to the aggregator, and the \emph{supplementary data flows} from external sources. The core flows are necessary for the VPP to function, while the supplementary flows provide context based on which the aggregator makes decisions. 

For the aggregator, the individual assets are expendable and interchangeable, as long as the capacity is sufficient, much like a classical employer-worker relationship. By forming a union, the individual asset-owners are able to leverage their combined control over the assets to influence the data flow, processing and storage within the VPP.

From this frame we have derived a set of principles for data governance models within VPPs: 

\begin{LaTeXdescription}

  \item [Collective bargaining] 
  The asset-owners should be able to bargain collectively about the conditions and purposes of the data flows. This includes which supplementary data flows to include and how to utilize them.
  
  \item [Representation] 
  The asset-owners should be represented in a central organizational governing body, which is in charge of defining and overseeing the data principles.

  \item [Disruptive measures] 
  The asset-owners should be able to limit or halt core data flow, allowing them to assert power and symmetry in the relation with the aggregator.

  \item [Accountability] 
  Transparency measures should be put in place to ensure the asset-owners ability to audit the data usage performed by the aggregator, in order to detect misuse and assign accountability

\end{LaTeXdescription}

Finally we have presented the issue in relation to data-oriented trade unions and the idea of "data commons", which both have similar traits to the data flows within VPPs.


\printbibliography

\end{document}